\documentclass[twocolumn,showpacs,preprintnumbers,amsmath,amssymb,nofootinbib]{revtex4-1}

\usepackage{graphicx}
\usepackage{dcolumn}
\usepackage{bm}
\usepackage{verbatim} 
\usepackage{epsfig}
\usepackage{amsmath}
\usepackage{amsthm}
\usepackage{subfigure}

\begin{document}

\date{\today}

\title{EDM planning using ETEAPOT with a resurrected AGS Electron Analogue ring}
\author{Richard M Talman}
\affiliation{
Laboratory of Elementary Particle Physics,
Cornell University, Ithaca, NY, USA}
   
\author{John D Talman}
\affiliation{
UAL Consultants, Ithaca, NY, USA}

\begin{abstract}
There has been much recent interest in directly measuring 
the electric dipole moments (EDM) of the proton and the electron, 
because of their possible importance in the present day observed
matter/antimatter imbalance in the universe. Such a measurement 
will require storing a polarized beam of ``frozen spin'' particles, 
15\,MeV electrons or 230\,MeV protons, in an all-electric 
storage ring. Only one such relativistic electric accelerator has 
ever been built---the 10\,MeV ``Electron Analogue'' ring at Brookhaven 
National Laboratory in 1954; it can also be referred to as 
the ``AGS Analogue'' ring to make clear it was a prototype for 
the AGS proton ring under construction at that time at BNL.
(Its purpose was to investigate nonlinear resonances as well as
passage through ``transition'' with the newly-invented alternating 
gradient proton ring design.) 
By chance this electron ring, long since dismantled and its 
engineering drawings disappeared, would have been appropriate both for 
measuring the electron EDM and to serve as an inexpensive prototype 
for the arguably more promising, but ten times more expensive, proton 
EDM measurement.

Today it is cheaper yet to ``resurrect'' the Electron Analogue 
ring by simulating its performance computationally. This is one purpose 
for the present paper. Most existing accelerator simulation codes cannot 
be used for this purpose because they implicitly assume magnetic bending. 
The new UAL/ETEAPOT code, described in detail in an accompanying paper, has 
been developed for modeling storage ring performance, including spin
evolution, in electric rings. Illustrating its use, comparing its 
predictions with the old observations, and describing new expectations 
concerning spin evolution and code performance, are other goals of the 
paper. To set up some of these calculations has required a kind of 
``archeological physics'' to reconstitute the detailed Electron Analogue 
lattice design 
from a 1991 retrospective report by Plotkin\cite{Plotkin} as well as
unpublished notes of Courant\cite{Courant} describing machine studies 
performed in 1954-55.

This paper describes the practical application of the ETEAPOT code and 
provides sample results, with emphasis on emulating lattice optics
in the AGS Analogue ring for comparison with the historical machine
studies and to predict the electron spin evolution they would have 
measured if they had had polarized electrons and electron polarimetry. 

Of greater present day interest is the performance to be expected for
a proton storage ring experiment. To exhibit the ETEAPOT code performance 
and confirm its symplecticity, results are also given for 30 million turn 
proton spin tracking in an all-electric lattice that would be appropriate 
for a present day measurement of the proton EDM.

The accompanying paper ``ETEAPOT: symplectic 
orbit/spin tracking code for all-electric storage rings''\cite{ETEAPOT1} 
documents in detail the theoretical formulation implemented in ETEAPOT, 
which is a new module in the Unified Accelerator Libraries
(UAL) environment\cite{UAL}.
\end{abstract}


\maketitle


\section{Introduction}

\paragraph{\bf Motivation for electric storage ring ``traps'' for 
electrons or protons.}
The U.S. particle physics community has recently updated its vision of the future
and strategy for the next decade in a Particle Physics Project Prioritization 
Panel (P5) Report. One of the physics goals endorsed by P5 is measuring 
the EDM of fundamental particles (in particular proton, deuteron, neutron and 
electron).

Since Standard Model EDM predictions are much smaller than current experimental 
sensitivities, detection of any particle's non-zero EDM would 
signal discovery of New Physics. If of sufficient strength, such a source could 
support an explanation for the observed matter/antimatter asymmetry of our 
universe. A proton EDM collaboration\cite{pEDM-PRL} has 
proposed a storage ring proton EDM measurement at the unprecedented level of 
$10^{-29}e \cdot\,$cm, an advance by nearly 5 orders of magnitude beyond the 
current indirect bound obtained using Hg atoms.

The proposed EDM measurement is based on the accumulation of tiny 
``wrong-plane'' (i.e. ``right'' for EDM, ``wrong'' for MDM) 
spin precessions that can be accounted for only by a non-zero EDM. 
Polarized beams can be frozen (for example in longitudinally-polarized 
states) in storage rings containing 
appropriate combinations of electric and magnetic bending elements;
the relative amount depends on the magnetic dipole moment of the 
particles being stored. Only for a few particles,
of which the electron and the proton are the most important, can the spins be
frozen in purely electric rings. This is highly advantageous, since
such rings support beams circulating both clockwise and counter-clockwise, 
permitting measurements for which important systematic errors cancel. 
The electric dipole moment (EDM) of the proton is known to be so small 
that, in order to reduce systematic errors enough to measure it in a 
storage ring requires measuring differences between counter-circulating 
beams. This can be done sequentially or, to provide differential beam position
precision at the cost of beam-beam complications, with simultaneous
circulating beams. (Direct colliding beam-beam effects are calculated 
to be negligible.) 

A comparably important advantage of electric bending is that the 
absence of intentional magnetic fields will reduce the presence of 
unintentional (radial) magnetic field components, which 
are expected to be the dominant source of spurious precession, 
mimicking the EDM effect. 

To freeze the spin procession, conventional storage ring bending magnets 
are replaced with the corresponding electric elements. Though the circulating 
particles 
are constantly being attracted to the inner electrode, their centrifugal force 
causes them to circulate indefinitely in a more or less circular orbit. 
This establishes the storage ring as an electric ``trap'', albeit for an
intense moving bunch rather than for a few slow particles.

In a storage ring EDM measurement, bunches of longitudinally 
polarized protons will circulate for ``long'' time intervals 
such as 1000 seconds. Because of inevitable parameter spreads, individual particle 
spins will precess differently and, after a 
spin coherence time \emph{SCT}, the 
beam polarization will have been attenuated (due to decoherence) to a point where 
the EDM precession rate has become unmeasurably small. The wrong-plane polarization 
difference between early and late times, when ascribed to the torque of
the bend field acting on the electric dipole moment, will provide a measurement
of the proton EDM.

For measuring EDM's the accumulation and measurement of small effects 
requires analysis, mitigation and control of various systematic errors. 
Issues to be studied with ETEAPOT include:

Radial B-field: since the torque due to any residual radial B-field acting on the 
magnetic dipole moment (MDM) mimics the EDM effect, this is expected to be the 
dominant systematic error. 

Geometric phase: some EDM-mimicking precessions would average to zero except for 
the non-commutativity of 3D rotations.

Non-radial E-field, vertical quad misalignment, RF cavity misalignment etc.: these 
cause the closed orbit to deviate from design. The systematic errors they cause 
cancel to ``lowest order'' but higher order effects need to be investigated.

Polarimetry: realistic beam distributions need to be used to identify and reduce 
left-right asymmetry bias, which is another source of systematic error.

\paragraph{\bf Initial storage ring simulation tasks.}
Initial tasks for the ETEAPOT code are to simulate the performance 
of arbitrary electric storage rings. As with any storage ring, these 
tasks (which now include also spin tracking) are
\begin{enumerate}
\item
Evaluation of linearized lattice functions such as tunes, the Twiss, 
$\alpha$, $\beta$, and $\gamma$ functions, dispersion function, 
and spin tunes, as well as chromatic dependence of these functions, 
and their sensitivity to imperfections. 
\item
Short term tracking, for confirmation of the lattice functions,
for determining dynamic aperture, and for investigating the 
performance of analytically-derived spin decoherence compensation 
schemes.
\item
Guaranteed to be \emph{stable}, long term tracking. The EDM experiment
requires \emph{SCT} to be 1000\,s or longer. 
During this time every particle executes about $10^9$ betatron
oscillations. The code is required to exhibit 
negligible spurious growth or decay for this interval of time. 
\end{enumerate}
Compared to magnetic bending,
electric bending complicates these tasks. This requires the ETEAPOT 
treatment to differ from the well-established TEAPOT treatment. 
As implemented in UAL/TEAPOT, lattice function determinations use 
a truncated power series formalism that has not (as yet) been
established for electric rings. Translating the arbitrary order 
formalism from magnetic to electric elements is a huge task that 
has only just begun. To cover all three storage ring simulation
tasks on the time scale required for EDM planning we have, therefore, 
had to proceed along more than one track.

One track starts by implementing exact (and therefore exactly symplectic)
tracking in the inverse square law potential electrical field between 
spherical electrodes. The optimal electric field shape will differ from
this, however. This makes it necessary to introduce artificial 
\emph{virtual} quadrupoles in the interiors of bend elements, 
in order to model deviation of the actual electric field away 
from its idealized representation. \emph{Real} quadrupoles are
typically also present in the lattice, for example to alter the 
lattice focusing, or to control dispersion. 
Thin sextupoles (present for example to adjust chromaticities or
to compensate spin decoherence effects) and other thin multipole
elements are also allowed. Causing only ``kicks'', these thin 
elements, either real or virtual, also preserve symplecticity. 
Now complete, this track allows tasks (2) and (3) to be 
completed for arbitrary lattices.

Since (linearized) transfer matrices are not used,
the ETEAPOT code cannot be used to extract Twiss functions
directly, as is commonly done in conventional Courant-Snyder  
accelerator formalism. However, tunes and Twiss functions can 
be obtained (to amply satisfactory accuracy) using FFT and MIA 
(model-independent analysis). Now complete this code also 
provides transfer matrix determination which, in turn, allows
determination of all the lattice functions required
to complete task (1).

Implementation of the UAL truncated power series approach
for electric rings has just been begun. Its progress is contingent 
on obtaining government funding\cite{BNL-Cornell-SMU}. 

After comparing ETEAPOT simulation results to results measured
on the AGS Analogue ring this paper briefly describes simulation of
spin evolution in a proton EDM storage ring.

\section{The AGS Electron Analogue ring}
Of the more than 100 relativistic accelerators ever built, only one
has used electric rather than magnetic bending. It was the AGS
Electron Analogue machine at BNL. Curiously it was also 
the first ring ever to use alternating gradient (AG) 
focusing. (The Cornell 1.1\,GeV alternating gradient
electron ring was commissioned at more
or less the same time and the BNL AGS ring itself somewhat later.) 

Along with using \emph{electrons} instead of \emph{protons}, and limited by
achievably high electric field, cost minimization of the AGS Analogue
led to the choice of 10\,MeV maximum electron energy, 4.7\,m bending 
radius, 6.8\,MHz RF frequency, and 600\,V RF voltage. 
These optimization 
considerations are very much the same as 
will be used to fix the parameters of a frozen spin proton ring
(which is tentatively expected to have a bend radius of 
about 50\,m)\cite{pEDM-PRL}. The UAL/ETEAPOT code 
(documented in the accompanying paper) was developed with this
application in mind. The present paper gives initial results.

Starting from fragmentary documentation, the paper starts
by reverse engineering the AGS Analogue lattice and producing 
a lattice description file 
{\tt E$\underline{\ }$AGS$\underline{\ }$Analogue.sxf},
in the format needed for processing in the Unified Accelerator
Libraries (UAL) environment. Results obtained using ETEAPOT are 
then compared with measurements performed on the ring at BNL in 
1954-55.

By chance, the magic kinetic energy for freezing electrons,
which is 15\,Mev, is not very different from the 10\,MeV
of the AGS Analogue ring. So that ring could have been used
to measure the electron EDM just by increasing the electric
bend field by a factor of 1.5. Morse\cite{Morse} and others have
suggested building such a ring for this purpose.
The ETEAPOT code can therefore be used to to simulate an
electron EDM measurement using a ring whose successful performance 
as a storage ring is all but guaranteed by the
successful operation of an ``identical'' ring in 1955.
15\,MeV is a very convenient electron energy and high quality 
electron source would be available, for example as described by 
Bazarov\cite{Bazarov}.

The present paper describes our ``resurrection'' of  AGS Analogue
ring from historic BNL documentation, and simulates its performance with 
codes intended for the proton EDM experiment.

A quite superficial (day long) search of the BNL library and the 
Accelerator-Collider report library found one quite extensive report,
produced retrospectively in 1991 by Martin Plotkin\cite{Plotkin},
and a few ancient reports describing machine studies results.
Also a report privately communicated from Ernest 
Courant\cite{Courant} contains experimental data to be 
simulated in this paper.

Before starting on this project, one of the authors, RT, benefited
from three brief but valuable meetings with Ernest Courant,
perhaps the father, or at least one of the parents, of the Electron 
Analogue ring, inquiring about his recollections concerning the
dynamics and performance of the ring. This contributed to our
reconstruction. Bill Morse\cite{Morse-EC} reports having had a similar 
conversation with Courant a few years earlier. Bill recalls asking 
Ernest whether, back in 1953, he (Courant) understood the difference 
between electric and magnetic focusing. Ernest's replied ``of course'' in 
his always kindly, but in this case somewhat exasperated, tone of voice. 

\begin{figure*}[ht]
\centering
\includegraphics[scale=0.42]{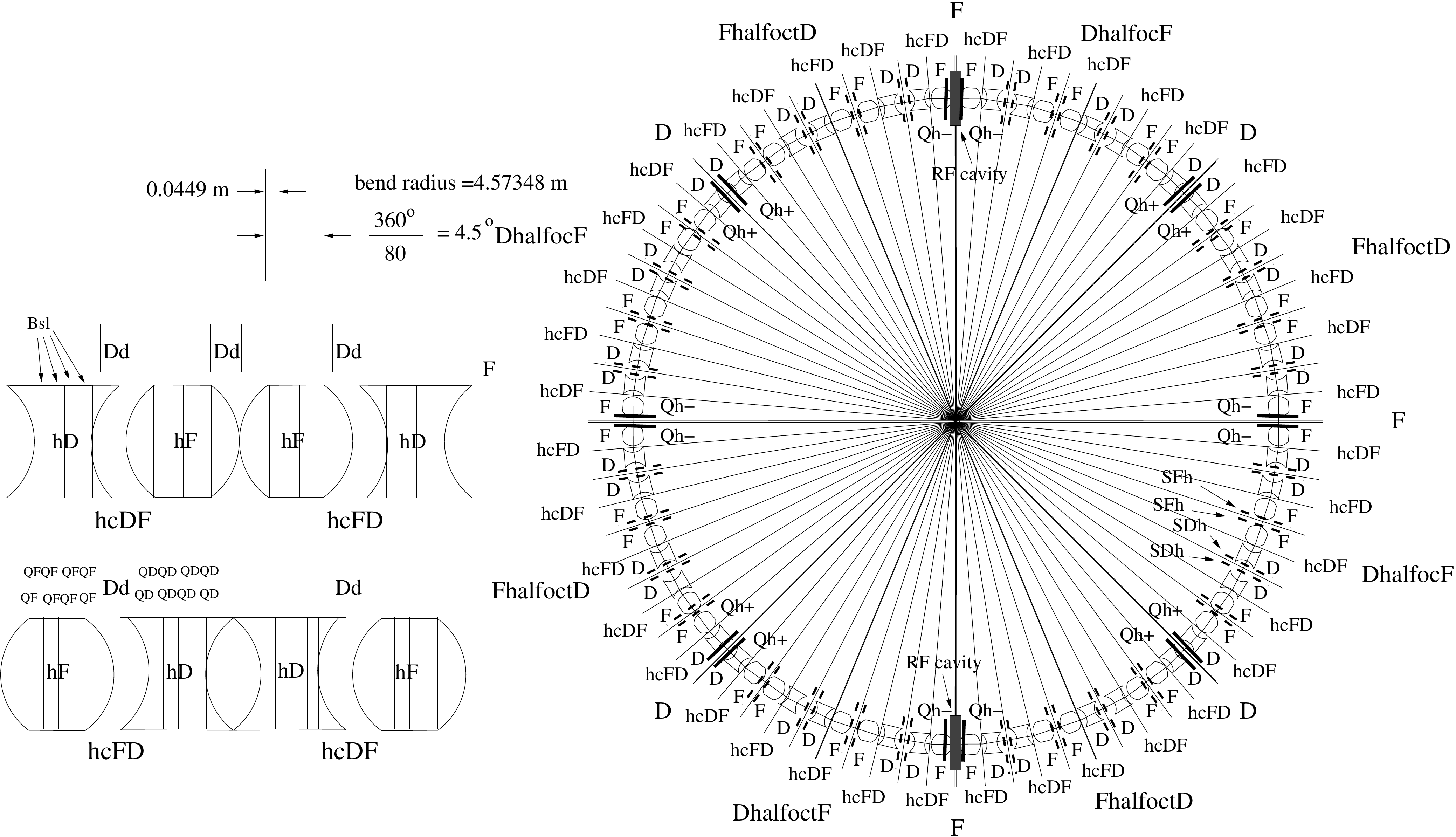}
\caption{\label{fig:AGS_Analogue}
The 1955 AGS-Analogue lattice as reverse engineered from available
documentation---mainly the 1953 proposal letter from BNL Director Haworth to
AEC Director of Research Johnson. Individual ``lenses'' (as they were
referred to in the original documentation; in modern terminology they would 
be referred to as ``combined function bends'') are shown on the left;
vertical lines indicate locations of thin quadrupoles present in the lattice
description to represent the focusing effect of the electric lens elements;
Elements are identified by \emph{ad hoc} labels assigned in the reconstruction. 
They are not obtained from the original lattice design, but are
needed for modern day lattice description files. Lattice description files
are avaiable in various formats: {\tt E\_AGS\_Analogue2.xml} gives the
pristine design with all parameters in algebraic (or function of algebraic)
form. {\tt E\_AGS\_Analogue2.adxf} is the same, but all parameters values, 
though ideal, converted to numerical values. {\tt E\_AGS\_Analogue2.sxf} 
contains the fully-instantiated lattice description (with parameters
from otherwise identical elements allowed to be individualized). Most UAL
lattice files are available in all these forms.}
\end{figure*}

\paragraph{\bf Historical BNL documents.}
The ``Conceptual Design Report'' for the AGS Analogue electron ring was a
four page letter, dated August 21, 1953, from BNL Director Haworth to 
the A.E.C. (predecessor of D.O.E.) 
Director of Research Johnson, applying for funding. The letter
is reproduced in its entirety in Figure~\ref{fig:HaworthLetter} in the
appendix. As brief as it is, this letter along with hints from Plotkin 
and Courant, includes everything needed to reconstruct the ring, 
as shown in Figure~\ref{fig:AGS_Analogue}. By 1955 the ring had been
approved, built, and commissioned, and had achieved its intended purpose.

Following the Haworth letter in the appendix are
two other especially informative figures from the paper by Plotkin. 
Figure~\ref{fig:AGS-AnalogueTanks} is especially useful for visualizing 
the physical layout of the AGS Analogue ring and its vacuum system. 
Figure~\ref{fig:PlotkinCourantData} 
is a more polished version of Figure~\ref{fig:QuadStrVsTunes}
which plots tune scans actually performed on 
the AGS-Analogue ring and reported by Ernest Courant in a July 28, 1955 
BNL technical report\cite{Courant}. 

In these plots, points of observed beam loss 
and observed beam disruption in the $(Q_x,Q_y)$ tune plane are correlated 
with expected resonances. Beam loss occurs on integer resonances, beam
disruption occurs on half integer resonances. The Courant report on, 
and analysis of, data collected in machine studies less than two years 
after the submission of the ring funding proposal, would 
certainly deserve an A+ 
grade by modern machine studies grading standards. 

The axes are voltages (proportional to focusing strengths)
applied to the tune-adjusting quadrupole families.
Short heavy lines indicate regions with no beam
survival (presumably due to integer resonance). 
Dots indicate points reported by Courant as
``showing the characteristic double envelope of the oscilloscope
pattern, sometimes accompanied by beam loss''. (These were presumably due
to ``very narrow'' 1/2 integer resonance).
The nominal central tunes values 
are $(Q_x,Q_y)=(6.5,6.5)$.  Stop bands due
to the eightfold lattice symmetry are also shown.

\section{Current day simulation of 1955 machine studies tune plane scan}
\begin{figure*}[ht]
\centering
\includegraphics[scale=0.45, angle=-90]{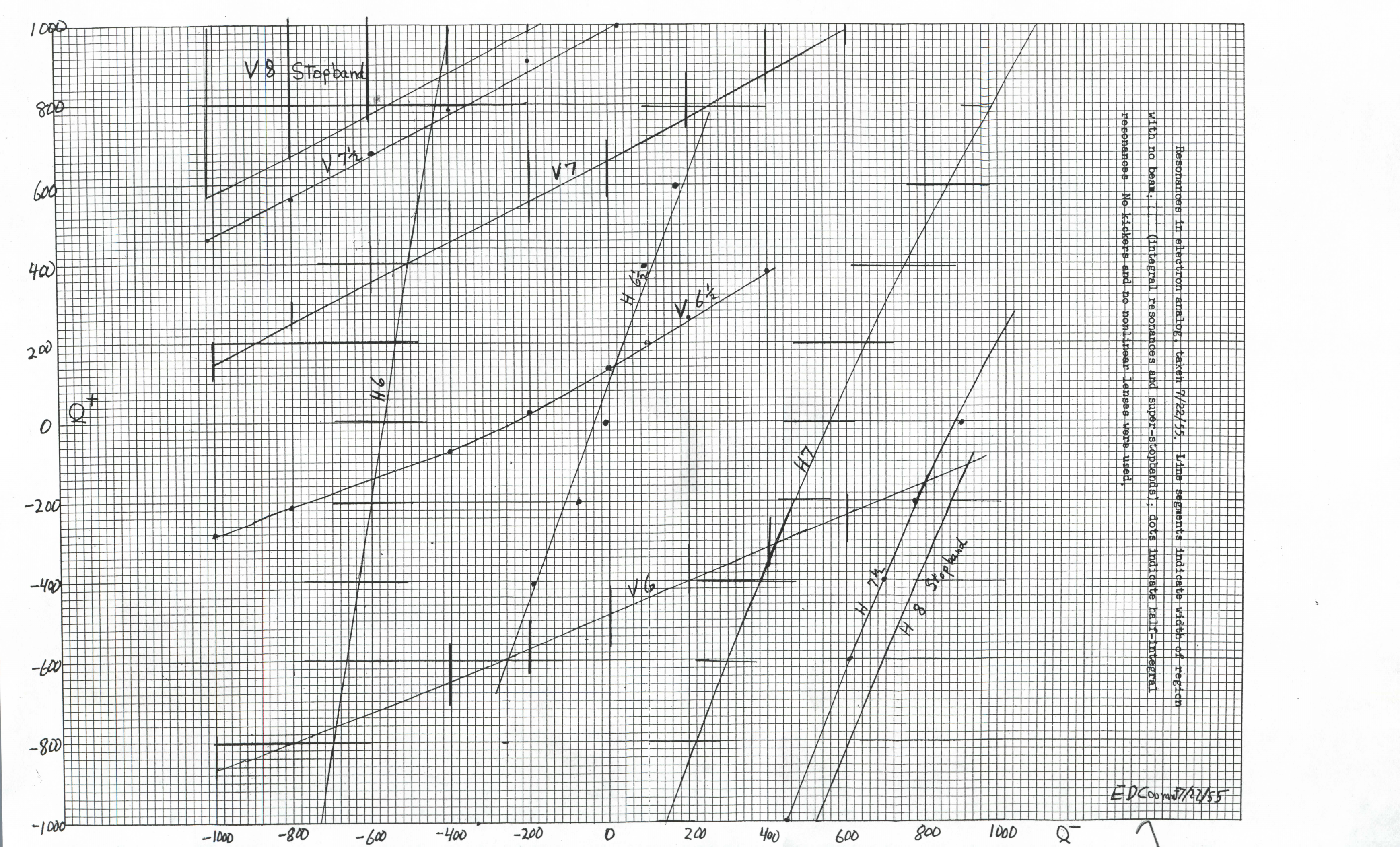}
\caption{\label{fig:CourantData}Tune plane resonance diagram measured 
during machine studies at the AGS Analogue ring and reported by 
Courant\cite{Courant}. A more polished and complete version of
this plot, copied from Plotkin\cite{Plotkin} is shown in 
Figure~\ref{fig:PlotkinCourantData} but with minor re-labelings.}
\end{figure*}

\begin{figure*}[ht]
\centering
\includegraphics[scale=1.0]{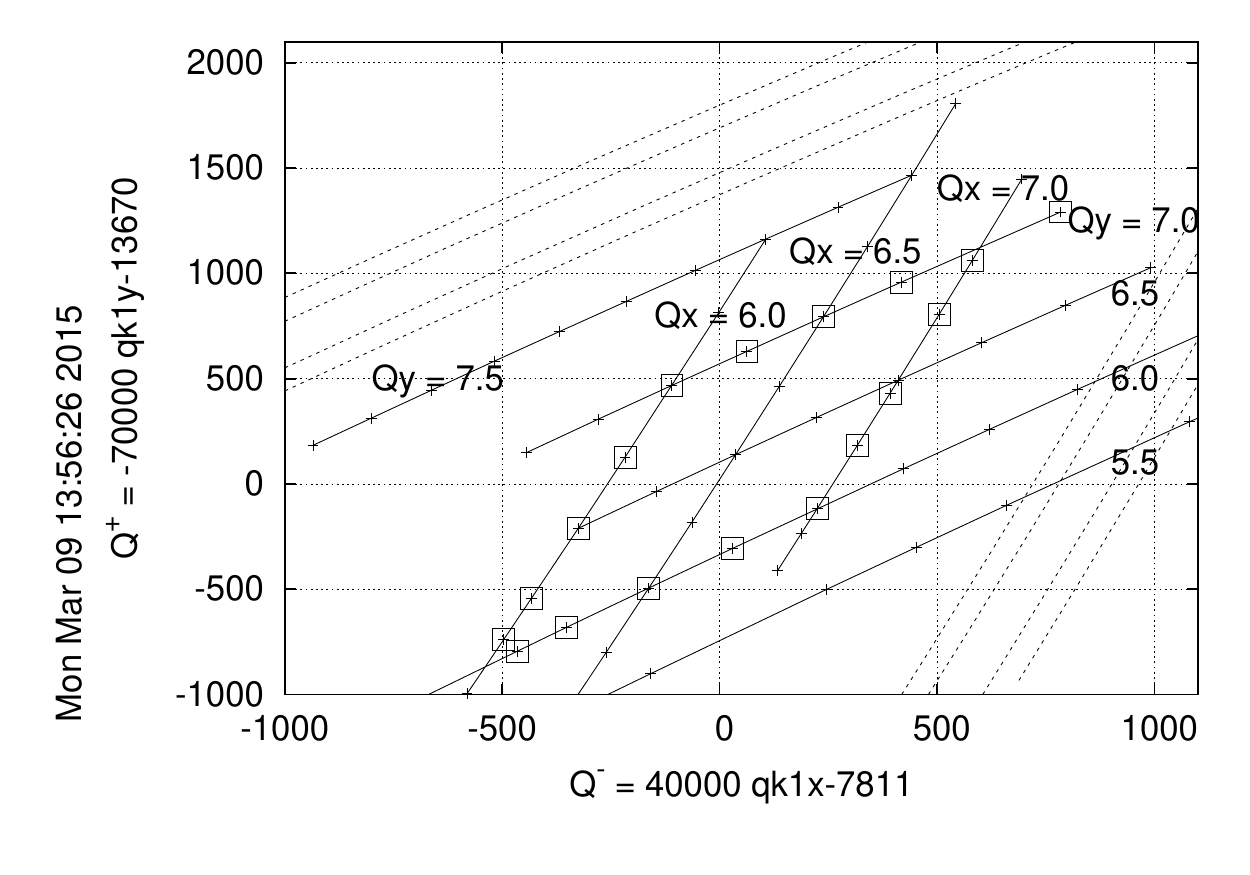}
\caption{\label{fig:QuadStrVsTunes}Tune plane resonance diagram
as calculated by TEAPOT, with bend elements treated as magnetic.
There is good qualitative and quantitative agreement with the
Courant data shown in Figure~\ref{fig:CourantData}.}
\end{figure*}

Our AGS Analogue lattice reconstruction is shown in Figure~\ref{fig:AGS_Analogue}.
The Courant tune plane plot is shown in Figure~\ref{fig:CourantData}. It is
to be compared with a similar plot, 
simulated by TEAPOT and shown in Figure~\ref{fig:QuadStrVsTunes}. As far as
we know this code and this lattice representation are completely equivalent 
Courant's model and analysis in 1955.

In the TEAPOT tune plane plot, boxes indicate points on integer 
resonance boundary curves of the stable diamond 
centered on nominal tune values $(Q_x,Q_y)=(6.5,6.5)$. 
Points lying on 1/2 integer resonance lines are indicated by dots. 
Superperiodicity (eightfold periodicity) causes
resonances with $Q_x=8$ or $Q_y=8$ indicated by broad blank 
bends bounded by narrowly-spaced lines. Courant refers to these
as ``stop bands''. Horizontal/vertical
axes $(Q^+,Q^-)$ are ``electrode voltages on the quadrupoles in 
odd/even-numbered tanks''. (For the meaning of ``tank'' see
Figure~\ref{fig:AGS-AnalogueTanks}.) The quadrupole strength 
coefficients for variable quadrupoles $(Q^+,Q^-)$
were determined empirically to match the central tunes. This
means that absolute focusing strength scales are not checked. 
Otherwise there are no significantly adjustable empirical lattice 
parameters. 

\begin{figure*}[ht]
\centering
\includegraphics[scale=0.3]{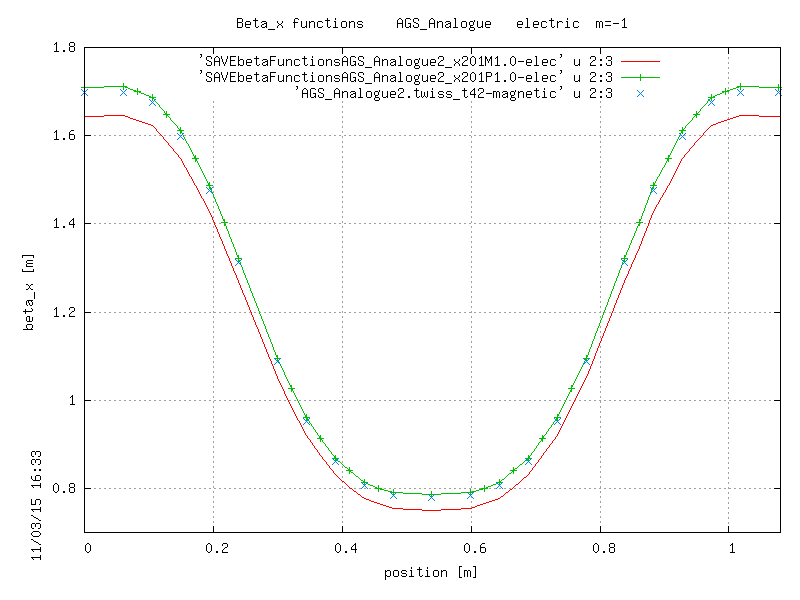}
\includegraphics[scale=0.3]{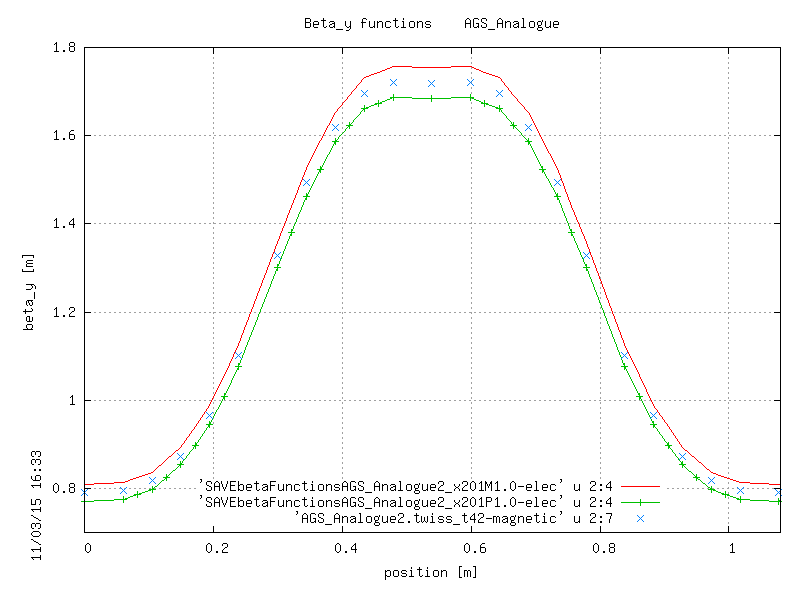}
\caption{\label{fig:ElecMagnCompare}Comparison for 
{\tt AGS\_Analogue2.sxf} lattice with bends treated as electric 
$m=1$ or $m=-1$ or magnetic. $\beta_x$ is plotted on the left,
$\beta_y$ on the right. As explained in the text, because of the strong 
focusing in the AGS Analogue lattice, switching from magnetic to
electric bending causes only minor changes to the lattice functions.}
\end{figure*}

The AGS Analogue ring provides only a coarse
test of ETEAPOT since, for strong focusing, the change from
electic and magnetic bending is quite minor.
This is illustrated in Figure~\ref{fig:ElecMagnCompare} which
shows that the $(Q_x,Q_y)$=(6.5,6.5) tunes in the AGS Analogue ring are 
high enough that the tune plane structure is quite insensitive to 
whether the bends are treated as magnetic or electric.

Treating the difference perturbatively, the effects of changing from
magnetic to electric bends are 
inversely proportional to $\beta$'s, which are both 6.5 in this 
case. For the eventual proton EDM ring the vertical tune has to be reduced from
$Q_y=6.5$ by at least a factor of ten which completely invalidates
any such perturbative estimation and requires radical adjustment
of quadrupole strengths. Reducing $Q_y$ from 6.5 to 2.25 has been
straightforward but, as the electric focusing became increasingly 
important, to decrease $Q_y$ further will require substantial design
effort.

\begin{figure*}[ht]
\centering
\includegraphics[scale=0.28]{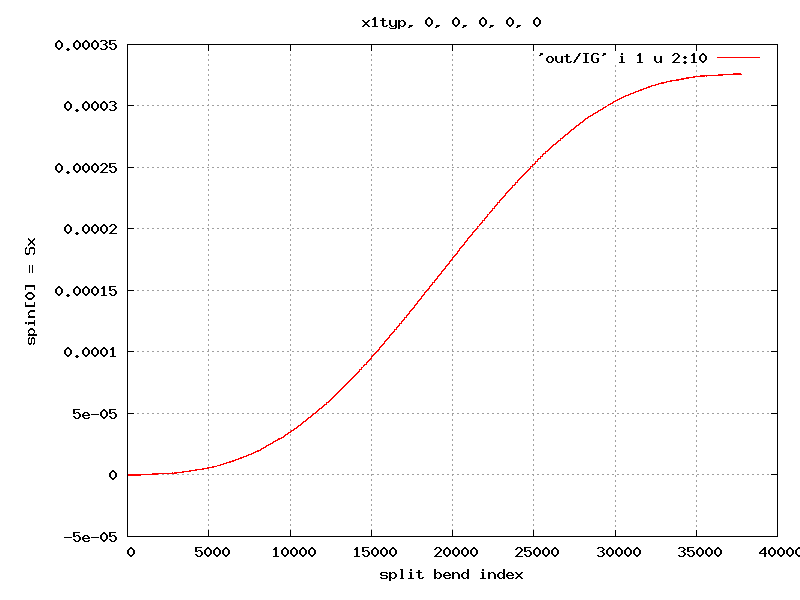}
\includegraphics[scale=0.28]{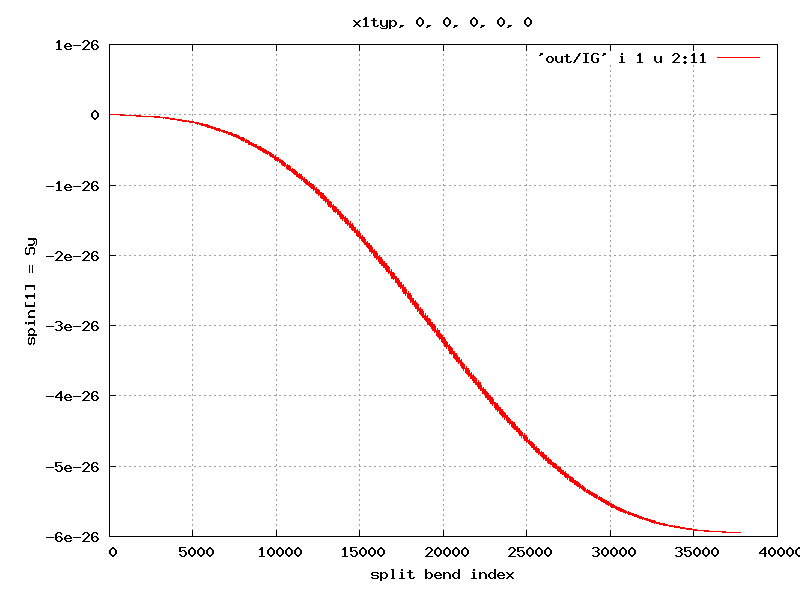}  \\
\includegraphics[scale=0.28]{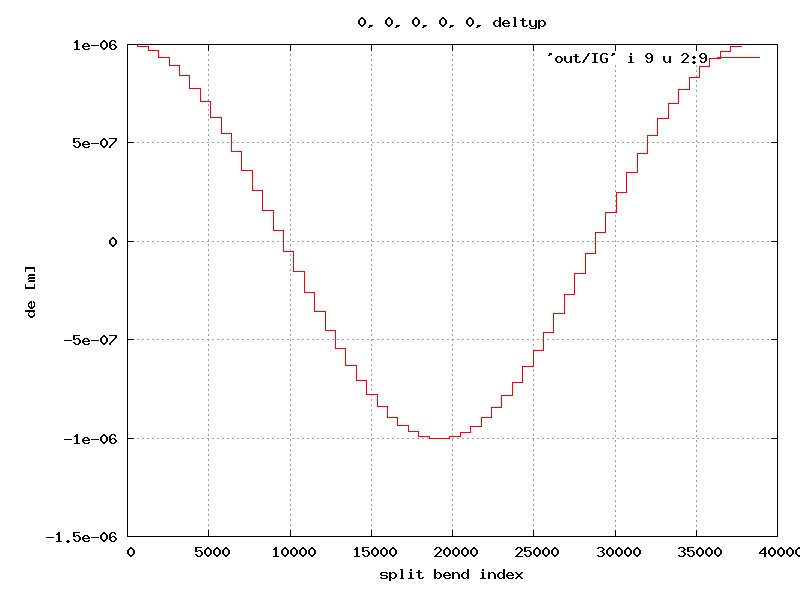}
\includegraphics[scale=0.28]{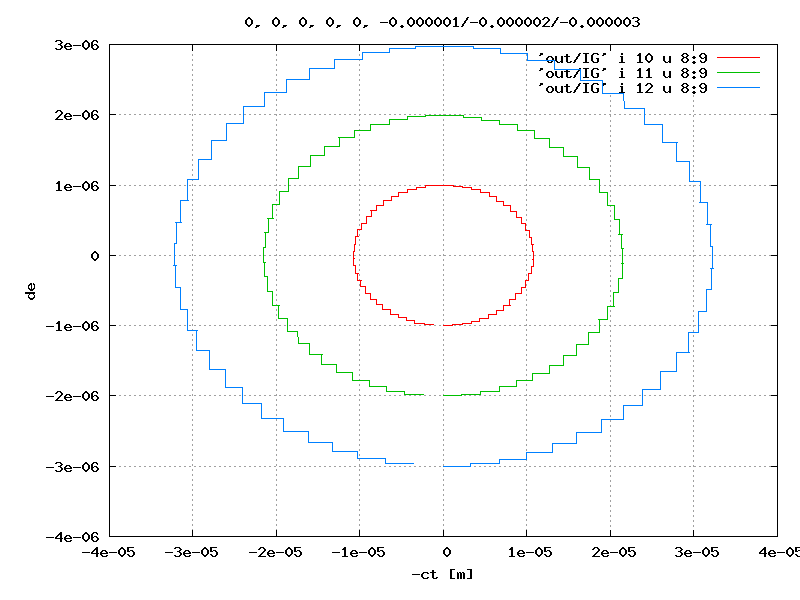}
\caption{\label{fig:6plus3Dplots}The upper llots show the $s_x$ and 
$s_y$ spin components of a single electron, with initial horizontal 
amplitude {\tt xtyp}=1e6\,m, in the AGS Analogue ring. (The 6D initial 
conditions are given above the plots.) The horizontal axis 
``split bend index'' increases by 1 at each bend, of which there are 640,
so about 59 turns are shown. The lower right figure shows longitudinal 
phase space evolution for three off-energy particles, with energy offset 
$de$ plotted against time offset $-ct$. One sees that almost exactly one 
revolution is completed during this time. The synchrotron tune is therefore 
approximately 1/59. The lower left plot plot shows the $de$ energy deviation 
during one synchrotron oscillation period.}
\end{figure*}

Spin evolution in the AGS Analogue ring is shown in 
Figure~\ref{fig:6plus3Dplots} and described in the caption.
From these results we are confident in out understanding of electric
rings and of our ability to simulate their performance using ETEAPOT.

\clearpage

\section{Long term tracking in a proton EDM storage ring trap}
Plots so far have simulated the evolution of 10\,MeV electrons in the AGS 
Analogue electron ring. These roughly indicate the performance to be expected in
a modern day electron EDM storage ring measurement. Of more immediate interest
is the expected behavior of 230\,MeV protons in a significantly larger, but 
still all-electric proton EDM storage ring.

\begin{figure*}[ht]
\centering
\includegraphics[scale=0.28]{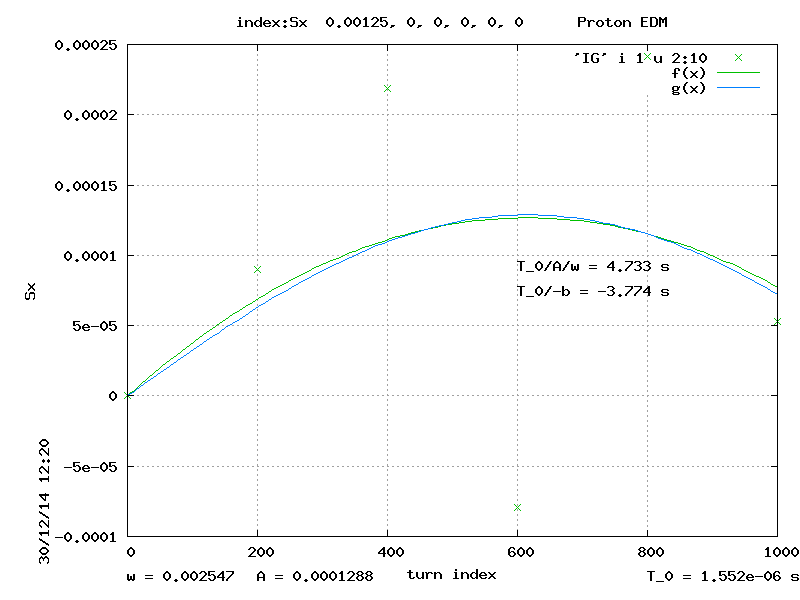}
\includegraphics[scale=0.28]{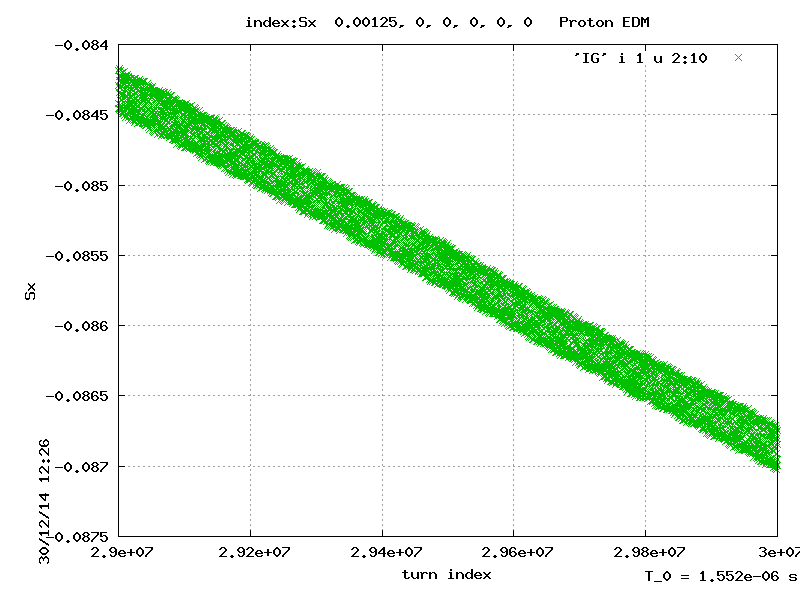}  \\
\includegraphics[scale=0.28]{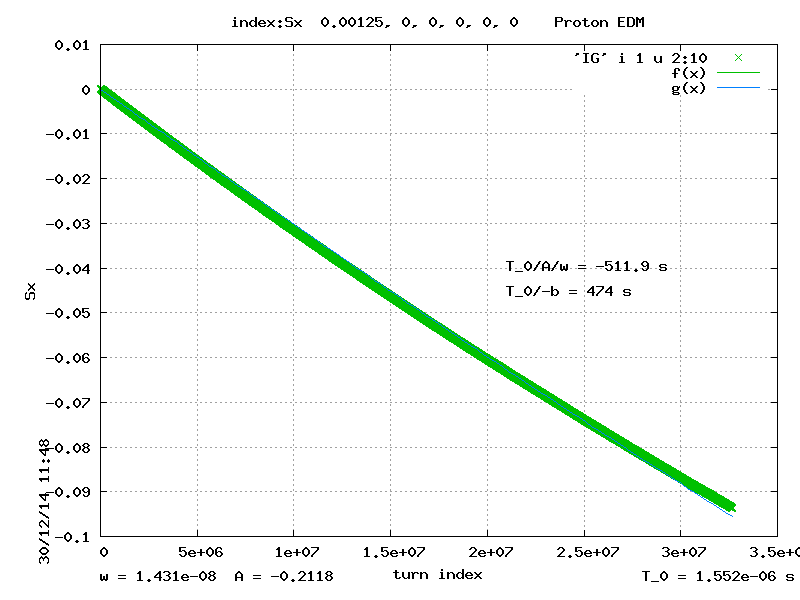}
\includegraphics[scale=0.28]{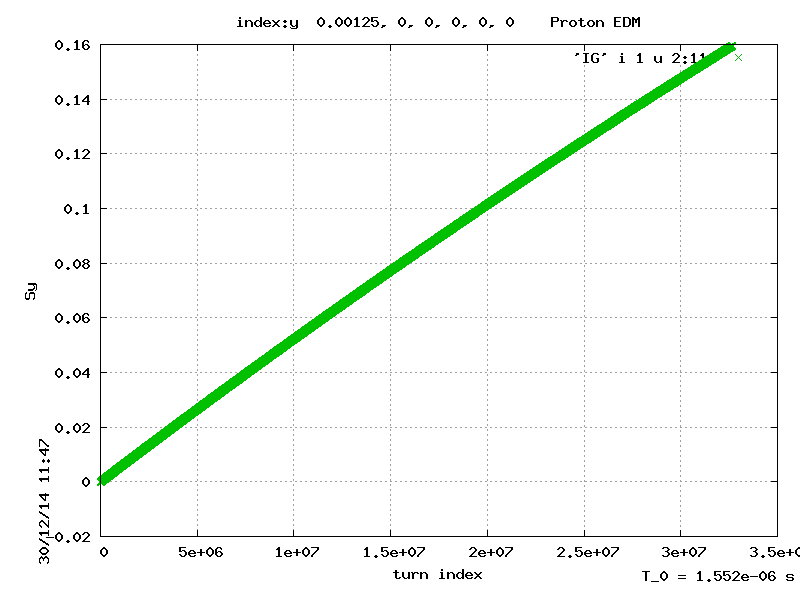}
\caption{\label{fig:plots}Evolution of spin components for 33 million turns. 
Because $S_z$=1 initially, both $S_x$ and $S_y$ vanish at $x$=0, where $x$ is 
turn number.
The upper left figure shows $S_x$ for the first 1000 turns, as fit by a sinusoid 
$A\sin(w x)$ where $x$ is turn number. The upper right figure shows $S_x$ for one 
million turns starting at $x=29\,$million. The lower left and upper right figures 
show the evolution of $S_x$ and $S_y$  for the full 33\,million turn run. Though
these motions are oscillatory the oscillation amplitudes are less than the 
line width. The lattice file is {\tt E\_pEDM-rtr1-Mobius.RF.sxf}.}
\end{figure*}

A single particle tracking example, taken from a recent proton EDM 
study, is illustrated in Figure~\ref{fig:plots}. 
The particle orbit and spin components are tracked around a prototype 
proton EDM storage ring for 33 million turns using ETEAPOT. As stated
earlier the tracking is exact and no artificial ``symplectification'' is 
applied. Any spurious damping or anti-damping of the spatial orbit over 
the full run is less than ten percent over the 33 million turns.
The (unit-magnitude) spin vector {\bf S} is initially purely tangential 
(forward) and the vector magnitude never changes from its initial value of 1. 

It is characteristic of spin evolution for the transverse spin components
to change over narrow ranges in synchronism with betatron oscillation
and larger but still small, ranges in synchronism with synchrotron oscillation.
The upper left graphs of Figure~\ref{fig:plots} show this, along with
a sinusoidal fit with the parameters shown. These precessions are
not expected to contribute significantly to spin decoherence. 
But any systematic growth over millions of turns will eventually lead to 
beam decoherence and limited spin coherence time. 

An important computational task in planning to measure the proton EDM is to 
determine the spin coherence time $SCT$ of the circulating beam. This is the
time after which inevitable spreads in beam parameters will have attenuated 
the beam polarization significantly due to decoherence in the spin propagation. 
As long as both $S_x$ and $S_y$ magnitudes remain small, {\bf S} remains in the 
forward hemisphere, and decoherence is suppressed. 

One sees from the graphs of Figure~\ref{fig:plots}
that both $S_x$ and $S_y$ gradually deviate from zero. 
But, for the particular particle being tracked, these deviations are limited
to small values. 
In particular, since $S_x^2+S_y^2$ remains always much less than 1 in magnitude, 
$S_z$ therefore remains not much less than 1. So the beam polarization 
remains always in the forward hemisphere. Depolarization is therefore
unimportant for particles of amplitude comparable to, or smaller than, 
this particle's. The tracking is therefore consistent with the SCT value 
exceeding the 33 million turns shown. 

The lattice investigated in Figure~\ref{fig:plots} is a
``M\"obius lattice'' in which the horizontal and vertical betatron 
oscillations interchange on every turn. This strongly suppresses spin
depolarization because precession in horizontal and vertical oscillation
phases tend to cancel on a turn-by-turn basis. By fine tuning this 
cancellation it is anticipated that extremely long SCT values can be 
obtained. 

This will leave beam energy spread as the dominant source of
beam depolarization. Fringe fields at bend elements are one such 
significant source of spin precession slewing. In this respect
spin tracking is more sensitive than orbit tracking. 
Fringe fields are not very important in orbit calculations. 
Treating bends as ``hard edged'' mainly causes small tune shifts which 
are not very important, since both tunes are always set operationally 
using spectrum analysis.
But, because of the spin precession sensitivity, in ETEAPOT, fringe 
fields are treated as linear ramps of length comparable with the gap 
between the electrodes, and not by hard edges. In any case synchrotron 
oscillation averaging also strongly suppresses this fringe
field depolarization. This places further demands on longitudinal 
particle tracking, which is always delicates, because nonlinearity of 
synchrotron oscillations contributes to spin decoherence.

The particle revolution period is about one microsecond so the plots shown 
correspond to about thirty seconds of real clock time in the laboratory. 
In a laptop computer this computation takes a few hours per particle. 
The ratio of computation time to real laboratory time
for a single particle is in the range from one to ten thousand.
The present ETEAPOT single particle tracking approach is sufficient for
early design tasks. 

EDM storage ring beams will contain perhaps $10^8$ particles.
Adequately precise simulation of the EDM experiment will require
fewer than this, but at least thousands of particles to be tracked 
for significantly longer times than shown. This will require heavy 
parallelization of the tracking. With little particle-to-particle 
interaction, the code is easily parallelizable. For more advanced tasks, 
such as investigating polarimeter biases or emittance growth caused 
intrabeam scattering, an efficient and scalable map-based 
computational approach is under consideration.

\appendix

\begin{figure*}[ht]
\centering
\includegraphics[scale=1.0]{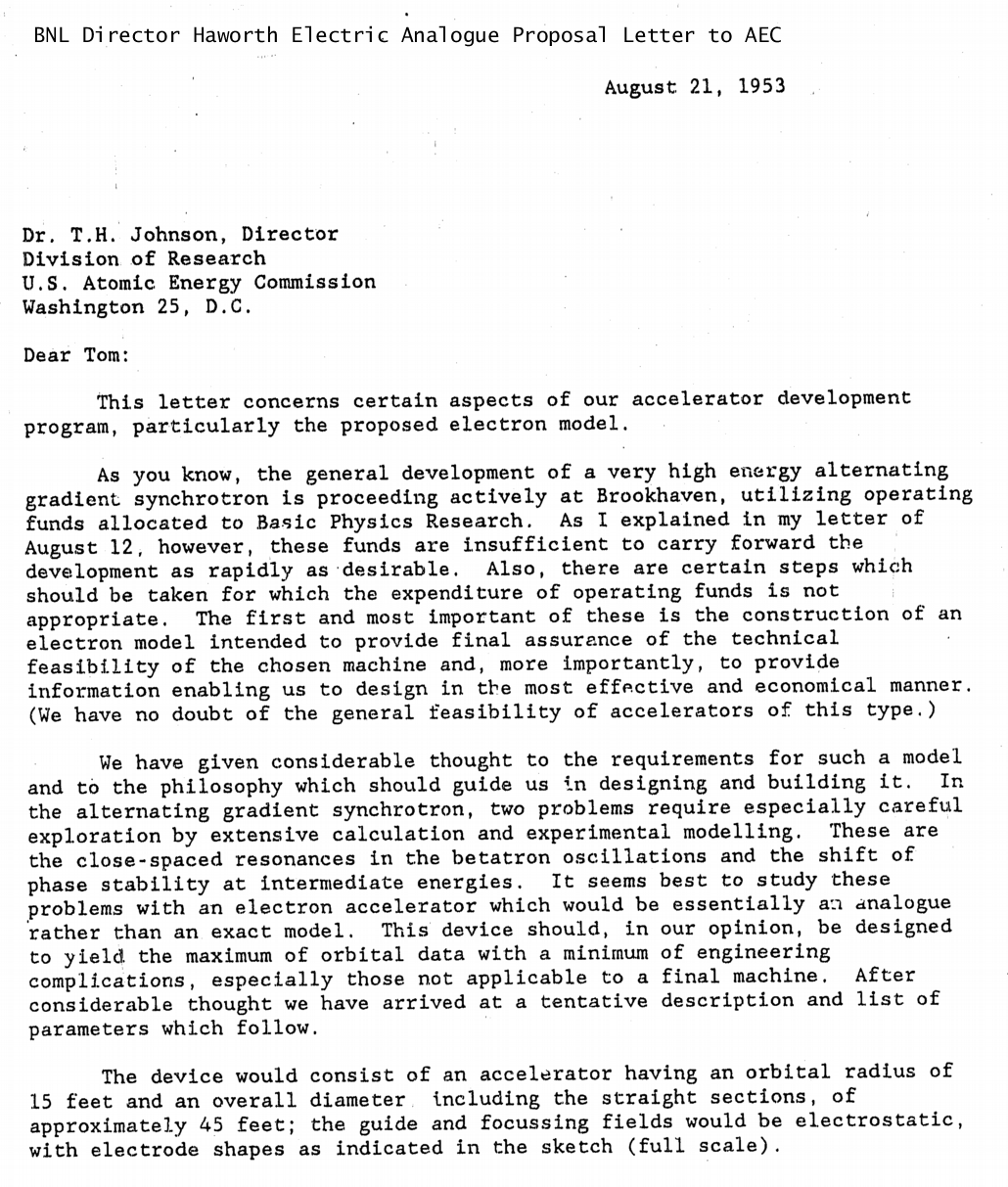}
\caption{\label{fig:HaworthLetter}This figure,
continued on the following three pages, is a copy of the original 
funding request letter from BNL for the AGS Analogue ring.}
\end{figure*}
\begin{figure*}[ht]
\centering
\includegraphics[scale=1.0]{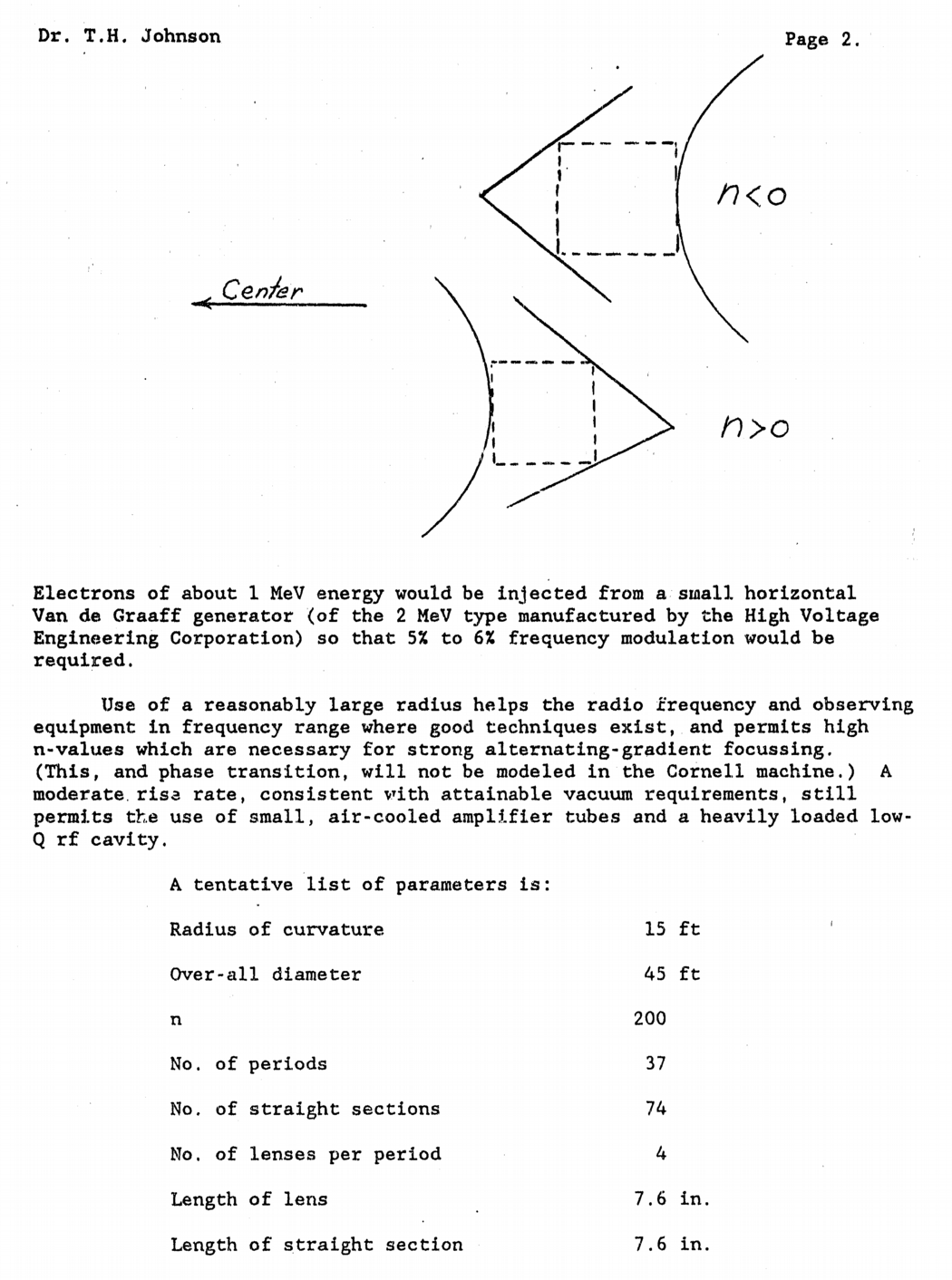}
\end{figure*}
\begin{figure*}[ht]
\centering
\includegraphics[scale=1.0]{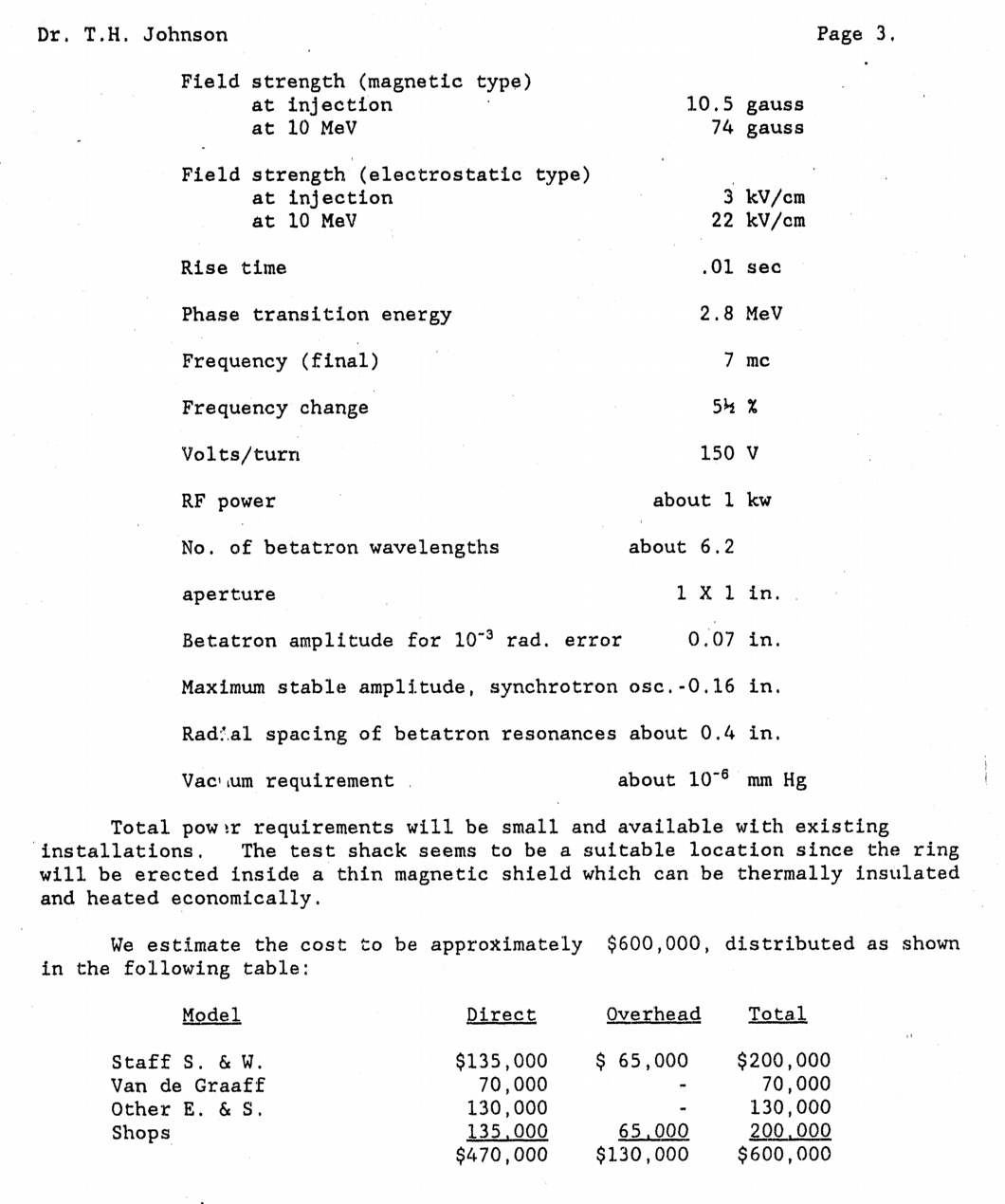}
\end{figure*}
\begin{figure*}[ht]
\centering
\includegraphics[scale=1.0]{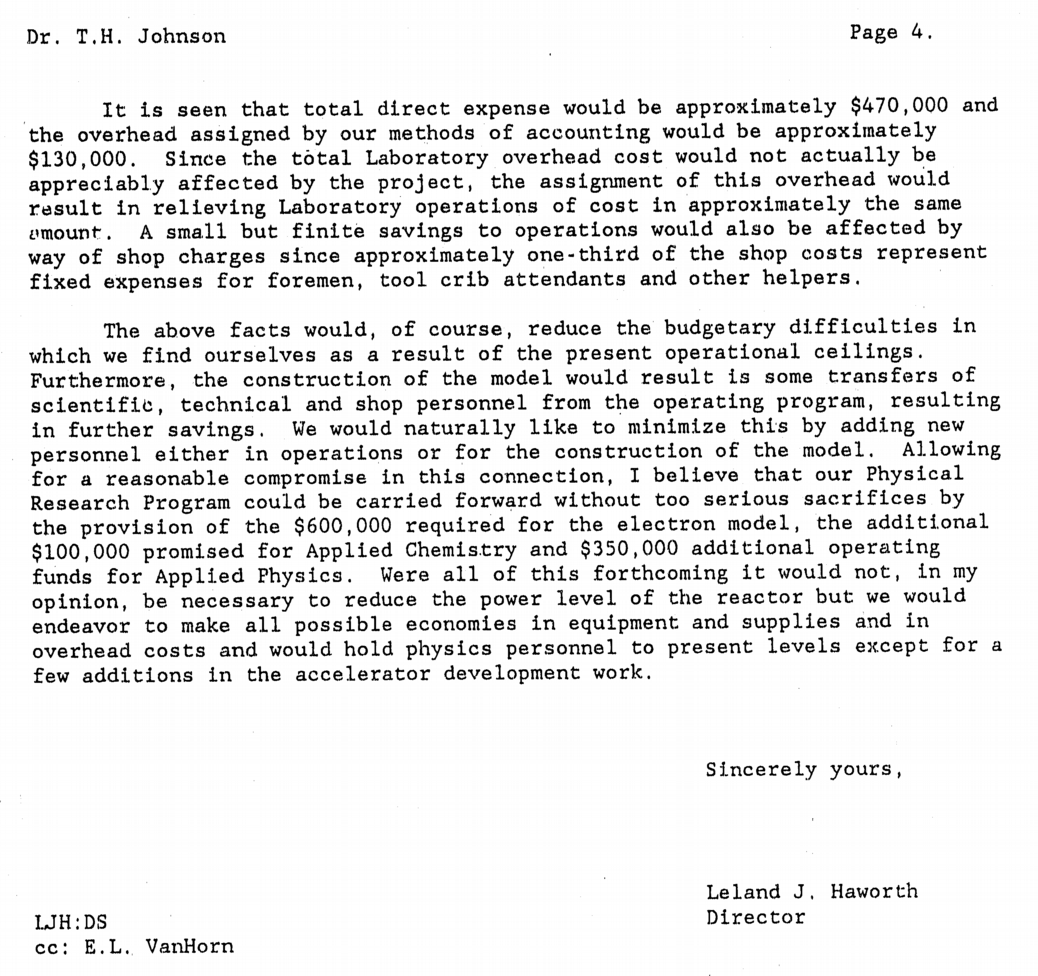}
\end{figure*}
\begin{figure*}[ht]
\centering
\includegraphics[scale=1.0]{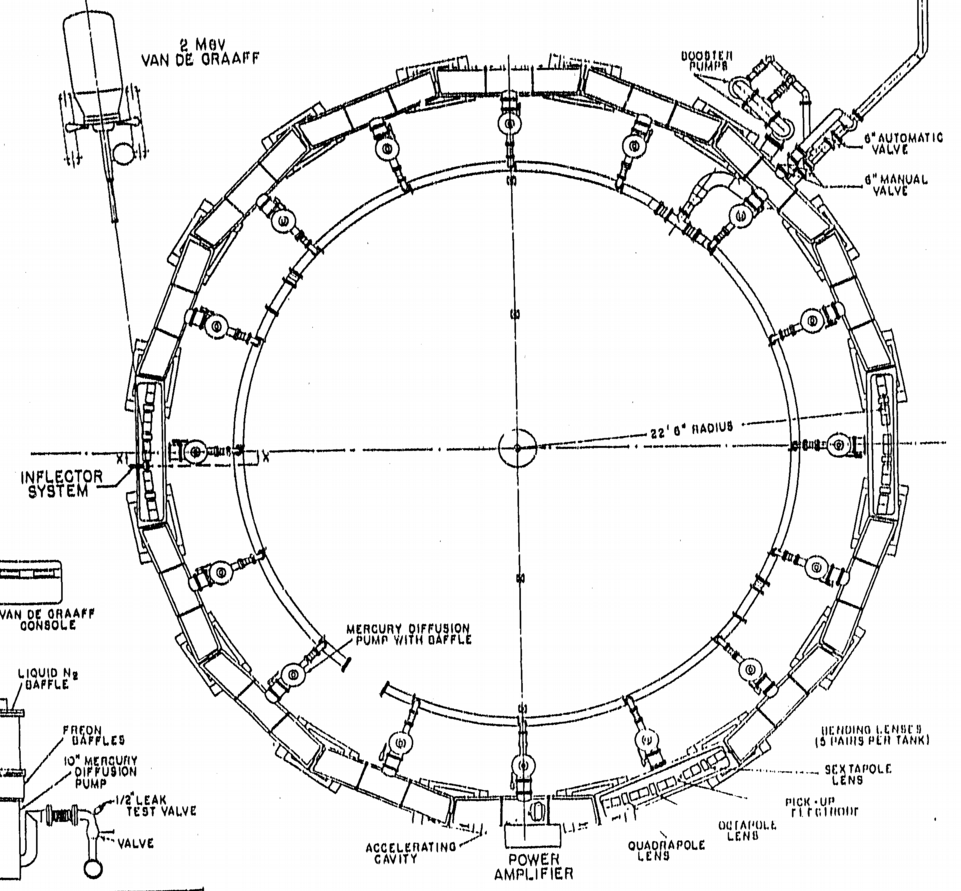}
\caption{\label{fig:AGS-AnalogueTanks}Electric ``leneses'' were contained,
10 to a tank, to the 16 vacuum ``tanks'' making up the full AGS Analogue
ring. Each tank therefore contained 2$\frac{1}{2}$ lattice cells, as
well as miscellaneous, special purpose quadrupoles, chromaticity-correcting
sextupoles and octupoles.}
\end{figure*}
\begin{figure*}[ht]
\centering
\includegraphics[scale=1.0]{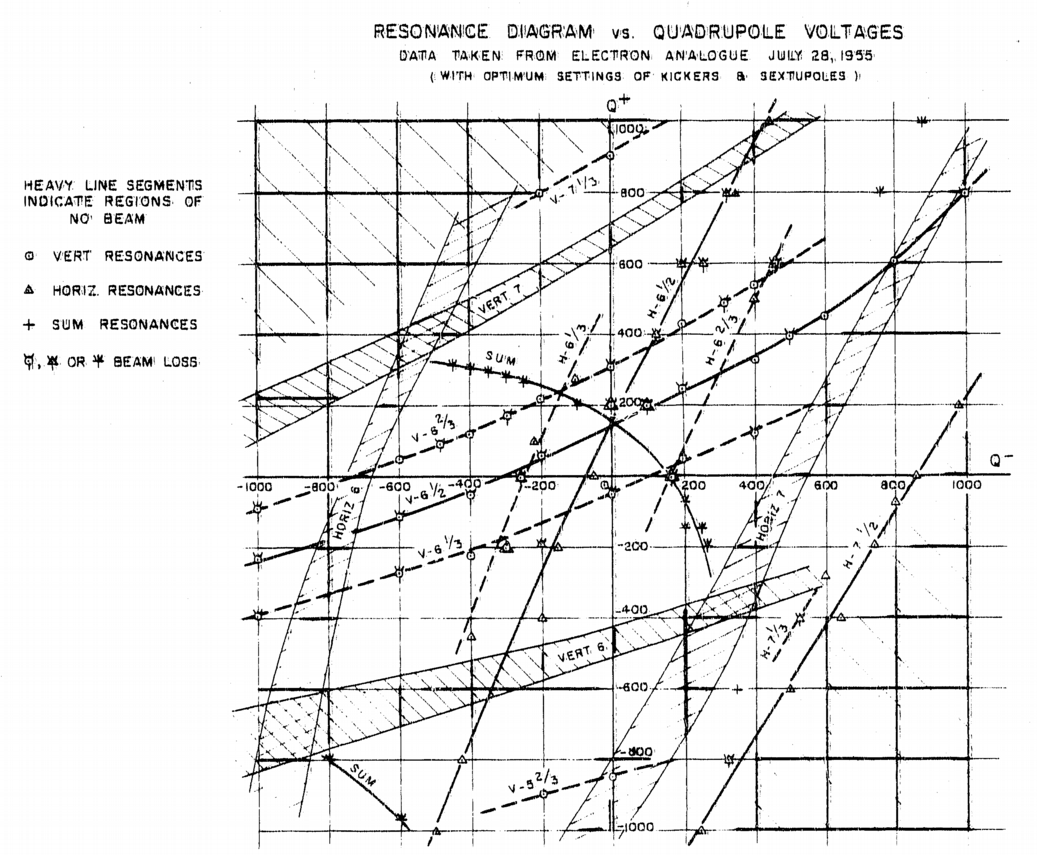}
\caption{\label{fig:PlotkinCourantData}This figure, from 
Plotkin\cite{Plotkin}, with little accompanying explanation, clearly 
represents the refined analysis of AGS Analogue machine studies data 
like that shown in Figure~\ref{fig:CourantData}.}
\end{figure*}

\clearpage

\end{document}